# Electrical Tunable Spintronic Neuron with Trainable Activation Function

Yue Xin[1], Kang Zhou[1], Xuanyao Fong[3], Yumeng Yang[1,2], Shenghua Gao[1], Zhifeng Zhu[1,2†]

[1]School of Information Science and Technology, ShanghaiTech University, Shanghai, China 201210

[2]Shanghai Engineering Research Center of Energy Efficient and Custom AI IC, Shanghai, China 201210

[3]Department of Electrical and Computer Engineering, National University of Singapore, Singapore 117576

Spintronic devices have been widely studied for the hardware realization of artificial neurons. The stochastic switching of magnetic tunnel junction driven by the spin torque is commonly used to produce the sigmoid activation function. However, the shape of the activation function in previous studies is fixed during the training of neural network. This restricts the updating of weights and results in a limited performance. In this work, we exploit the physics behind the spin torque induced magnetization switching to enable the dynamic change of the activation function during the training process. Specifically, the pulse width and magnetic anisotropy can be electrically controlled to change the slope of activation function, which enables a faster or slower change of output required by the backpropagation algorithm. This is also similar to the idea of batch normalization that is widely used in the machine learning. Thus, this work demonstrates that the algorithms are no longer limited to the software implementation. They can in fact be realized by the spintronic hardware using a single device. Finally, we show that the accuracy of hand-written digit recognition can be improved from 88% to 91.3% by using these trainable spintronic neurons without introducing additional energy consumption. Our proposals can stimulate the hardware realization of spintronic neural networks.

**I. Introduction**

The rapid development of deep learning calls for the hardware realization of neural networks with higher performance and lower energy consumption. CMOS implementations of artificial neurons and synapses occupy large silicon area and they are also energy inefficient due to the volatile nature [1, 2]. In contrast, spintronics has the intrinsic nonvolatility, based on which the spin-based neurons and synapses with very low energy consumption have been proposed [3-8]. In addition, the rich underlying physics enables them to realize complex functions using a simple device. As a result, they can be more compact than their CMOS counterparts. In particular, the stochastic magnetic tunnel junction (MTJ) has been proposed to produce different activation functions, e.g., sigmoid [5, 9-19], ReLU [17, 20], linear [18] or step [2]. However, the activation functions in these works are fixed in shape during the training of neural network. This limits the weight update during the network training. Intuitively, the performance of neural network can be greatly improved if the activation function is allowed to change during the training process. For example, in the beginning, one might need the output to change rapidly since the output and the correct value have large difference at this moment, whereas one requires it to changes slowly in the end. In this case, the activation function should have a sharp slope in the beginning and gradually becomes smoother. In addition, one of the most important algorithms used in the machine learning is the batch normalization [21], where the inputs are shifted and renormalized after each iteration. The idea behind the batch normalization is that during the training of neural network, the input distribution gradually deviates from the ideal one that is suitable for the training. This deviation can be remedied by shifting and renormalization of the inputs. As we will see below, the trainable activation function proposed in this work performs the similar role as the batch normalization, and our work points out the possibility of hardware realization of important algorithms by exploring the spintronic physics.

Following this idea, in this work, we exploit the physics of magnetization switching to realize the electrically controlled neuron with trainable activation function. The electrical controllability is necessary since the activation function is required to change dynamically during the training process. When the spin torque is used to switch the MTJ, we find that the slope of curve can be tuned by the pulse width, pulse amplitude or magnetic anisotropy, all of which can be electrically controlled. After that, the slope is used as an additional parameter in the neural network. Similar to the learning of weights, the updating of slope is guided by the gradient descent algorithm. However, they are different in the physical realization, i.e., the slope is a local parameter, which can be changed by applying electrical current on neurons. In contrast, the weights are nonlocal parameters. They denote the connection between different neurons, and can be realized using the crossbar resistive network. Thus, the introduction of activation function gradient in this work adds a new degree of freedom to improve the neural network, which has not been discussed in previous studies. Since the slope is obtained through the training process, we call it as the trainable activation function. The performance of this spintronic neuron is then evaluated. We first study the one without trainable activation function, which gives a recognition accuracy of 88%. Next, the trainable slope is included by controlling both the pulse width and magnetic anisotropy, resulting in an improved accuracy of 91.3%. In addition, the number of iterations required for the learning is greatly reduced.

**II. Results and Discussions**

The device studied in this work consists of the MTJ on the top of a ferroelectric (FE) layer [see Fig. 1(a)], which will be used to control the anisotropy of the free layer [22]. The widely used CoFeB is chosen as the free layer. A voltage ($V_{MTJ}$) or current pulse ($I_c$) is applied to switch the magnetization state via the spin-transfer torque (STT). The magnetization dynamics is calculated by solving the Landau-Lifshitz-Gilbert-Slonczewski (LLGS) equation [23-26] $d\mathbf{m}/dt = -\gamma \mathbf{m} \times \mathbf{H}_{eff} + \alpha \mathbf{m} \times d\mathbf{m}/dt - \gamma \hbar J_c/(2et_{FL}M_s)\mathbf{m} \times (\mathbf{m} \times \boldsymbol{\sigma}_{SOT})$ with the effective magnetic field, $\mathbf{H}_{eff}$, the damping constant, $\alpha = 0.0122$, the gyromagnetic ratio, $\gamma = 1.76 \times 10^{11}$ rad/(s·T), the reduced Planck constant, $\hbar = 6.58 \times 10^{-16}$ eV·s, the electron charge, $e = 1.6 \times 10^{-19}$ C, the free layer thickness, $t_{FL} = 1.3$ nm, the saturation magnetization, $M_s = 1.58$ T [27, 28] and the spin polarization, $\boldsymbol{\sigma}_{STT}$, whose direction depends on the polarity of $V_{MTJ}$, i.e., a positive and negative $V_{MTJ}$ results in $\boldsymbol{\sigma}_{STT}$ along $-\mathbf{z}$ and $+\mathbf{z}$, respectively. $\mathbf{H}_{eff}$ consists of the crystalline anisotropy field, $\mathbf{H}_{an}$, the demagnetizing field, $\mathbf{H}_{demag}$, the thermal field, $\mathbf{H}_{thermal}$, and the external field, $\mathbf{H}_{ext}$. $\mathbf{H}_{an} = 2(K_{bulk} + K_i/t_{FL})m_z\hat{\mathbf{z}}/M_S$ with the bulk anisotropy $K_{bulk} = 2.245 \times 10^5$ J/m$^3$, the interface anisotropy $K_i = 1.286 \times 10^{-3}$ J/m$^2$ [27]. We have verified that the $K_u$ is sufficient to overcome the shape anisotropy so that the equilibrium magnetization aligns in the **z** axis. $\mathbf{H}_{demag} = -(N_x m_x \hat{\mathbf{x}} + N_y m_y \hat{\mathbf{y}} + N_z m_z \hat{\mathbf{z}})$ with the demagnetizing tensors ($N_x$, $N_y$ and $N_z$) calculated based on the sample geometry (the rectangular sample with length and width of 15 nm and 15 nm, respectively). $\mathbf{H}_{\text{thermal}}=N_1(0,u)\hat{\mathbf{x}}+N_2(0,u)\hat{\mathbf{y}}+N_3(0,u)\hat{\mathbf{z}}$ is a random field where $N_i(0,u)$ is the normally distributed random variable with a zero mean and a standard deviation, $u=\sqrt{2k_\text{B}T\alpha/(V_{\text{FL}}M_\text{S}\gamma(1+\alpha^2)\Delta t)}$. $k_\text{B}=1.38\times10^{-23}$ J/K, $T=300$ K, $V_{\text{FL}}$, and $\Delta t=5$ ps are the Boltzmann constant, the temperature, the volume of free layer, and the simulation time step, respectively. Due to the thermal noise, the magnetization switches under $I_\text{c}$ with some probability, $P_{\text{sw}}$ [29-42], which is calculated by simulating the magnetization dynamics at each $I_\text{c}$ for 100 times (see Appendix A). Fig. 1(b) shows that $P_{\text{sw}}$ under different pulse widths can be well described by a sigmoid function in the form $f=1/(1+\exp(-k(I_\text{c}+c)))$, where $k$ and $c$ denotes the slope and shift of the curve (see Appendix B), respectively. As a result, controlling the pulse width enables the use of this MTJ as an electrically tunable neuron with sigmoid activation functions. To assess the performance of this spintronic neuron, we build a simple three-layer neural network to classify hand-written digits from the MNIST dataset [43], which consists of 50000 training samples and 10000 test samples. As shown in Fig. 1(c), the input layer has 784 neurons corresponding to the number of pixels in the picture. The hidden layer and output layers are configured as the sigmoid types, which have 25 and 10 neurons, respectively. We then train the neural network in the standard way. Firstly, the forward propagation is performed to compute the loss, $J=-[y\log(a_3)+(1-y)\log(1-a_3)]$, where $a_3$ is the output from the last layer and $y$ is the correct input value [i.e., $y=8$ in Fig. 1(c)]. Next, the backpropagation algorithm is used to update the weights ($\mathbf{w}_1$ and $\mathbf{w}_2$) so that the loss can be reduced. These two steps are then repeated until the loss is minimized. As shown in Fig. 1(d) as the filled triangles, the loss reduces monotonically as the training proceeds, indicating that the neural network learns correctly. Next, the test samples are used to evaluate the recognition accuracy of the network.

This can correctly reflect the performance of the neuron since these samples are new to the neural network. The filled circles in Fig. 1(d) show that the recognition accuracy reaches 88% after 1000 iterations.

During the fitting of $P_{sw}$ using the sigmoid function, we find that there are two additional parameters, i.e., $k$ and $c$. For a larger pulse width, the required switching current is reduced, which explains the shift of curves (i.e., a finite $c$). In addition, more electrons are injected into the magnetic free layer when the pulse width is larger, which leads to a stronger spin torque and thus, faster switching and sharper curves (i.e., larger $k$). It is therefore intuitive to think that the additional freedoms in tuning $k$ and $c$ can improve the performance of neural network. The idea behind this is similar to the batch normalization algorithm that is widely used in the machine learning. In the batch normalization, the inputs are shifted and renormalized after each iteration, which are performed in the software. In our proposed device, the shifting of inputs (i.e., $I_c$) is accomplished by $c$, and the multiplier $k$ scales the inputs, which is similar to the process of normalization. Different from the software implementation, the control of $k$ and $c$ can be realized in the spintronic device. Furthermore, since both $k$ and $c$ can be electrically tuned by changing the pulse width, they can be used as learned parameters for the neural network which are dynamically updated during the training process. It is worth noting that $k$ and $c$ are local parameters with respect to the neurons on which the current pulse is applied. This is different from the weights, which are nonlocal parameters that represent the connection between neurons in different layers. For the physical realization, the weights can be realized using the crossbar resistive array (see Appendix C for a detailed description). For the numerical representation, the dimension of weights is related to the topology of neural network. For example, $\mathbf{w}_1$ has a dimension of 784-by-25 since it connects the input layer (784 neurons) and the hidden layer (25 neurons). In contrast, $k_2$, which represents the slope of activation functions in the hidden layer, has the dimension of 25-by-1 that is identical to the dimension of local neurons. After adding $k$ and $c$ into the neural network, we have also derived their backpropagation equations (see Appendix D), following which the values of $k$ and $c$ are dynamically updated during the training process. Since the optimum $k$ and $c$ are obtained through training, we name the neuron proposed in this work as the one with trainable activation function. The performance of this neural network is shown in Fig. 1(d) as the empty circles, where a significant improvement in the accuracy (91.7%) is observed.

In the previous discussions, we have assumed that any values of $k$ and $c$ are allowed, which can only be realized in the software implementation. By tracing the changes of $k$ and $c$ during the training, we find that the ranges of $k_2$, $k_3$, $c_2$ and $c_3$ are −3.1 to 3.2, −3.6 to 3.6, −0.6 to 0.5 and −0.8 to 0.9, respectively [see Fig. 2(a)]. However, in the hardware implementation, one has to evaluate the allowed values provided by the device, which is determined by the underlying physics. We have learned in Fig. 1(b) that the change of pulse width leads to the different sets of $k$ and $c$, which has been summarized in Fig. 2(b). The relation between $k$ and pulse width shows a linear trend, and the $k_1$ and $k_2$ required by the backpropagation algorithm [cf. the blue symbols in Fig. 2(a)] can be satisfied by changing the pulse width. In contrast, the values of $c$ appear to be inconsistent between the one required by the algorithm [e.g., smaller than 1 µA as shown in in Fig. 2(a)] and that provided by the spintronic physics [e.g., larger than 10 µA as shown in in Fig. 2(b)]. Furthermore, it is important to notice that $k$ and $c$ are coupled together, e.g., the increase in pulse width reduces the required switching current as well as speeds up the magnetization switching. This produces a problem in the implementation of the trainable activation function. When one adjusts the pulse width to get the $k$ required by the backpropagation algorithm, the $c$ also changes to a new value $c_a$ due to the coupling between $k$ and $c$ that is determined by the spintronic physics. At the same time, similar to the update of $k$, the algorithm also requires $c$ to change to a value $c_b$, which is unlikely to be the same as $c_a$. This inconsistency between the algorithm and hardware physics would lead to the failure of training. We have tested this by setting up a simulation, in which the change of $k$ is guided by the backpropagation algorithm. At the same time, $c$ is changed to a new value following the relation between $k$ and $c$ [cf. Fig. 2(c)]. The simulation results show that the training fails in a few iterations (not shown here). This can be understood using the following illustrations. A small change in $k$ would lead to a large change in $c$ [change of $k$ from the circle to the triangle in Fig. 2(c)], resulting in a shifted activation function as shown in Fig. 2(d). Therefore, under the same input current ($I_{old}$), the slope becomes zero in the new curve. This indicates that the weights can never be changed, and thus the training cannot be proceeded.

One of the solutions is to decouple $k$ and $c$ by exploiting the device physics and to add a bias current so that $c$ is shifted to the value required by the algorithm. The implementation of this solution will lead to the improved accuracy as shown in Fig. 1(d). Alternatively, here we propose a solution that is easier to be implemented. Recall that the problem is raised by two facts, (1) $k$ and $c$ are coupled together, and (2) the inconsistency between the value of $c$ required by the algorithm and the one provided by the spintronics physics, a straightforward solution is to remove $c$ in the trainable activation function and only use the $k$. This would require an activation function where the slope can be changed without any shift in the curve. We will see below that this can be realized by exploiting the physics of magnetization switching.

Up to now, the activation function is only tuned by changing the pulse width. In fact, the properties of MTJ allows one to use other electrical means. For example, the effect of pulse amplitude is similar to the pulse width. Pulse with larger amplitude is shifted to the left, and the switching becomes sharper. In contrast, when the energy barrier ($E_B$) is reduced by modifying the $K_u$, only the left shift of curves is observed and there is no change in the slopes [see Fig. 3(a)]. Since $K_u$ determines the energy barrier of the system, smaller $K_u$ leads to a smaller switching current. However, since the input currents are identical, the speed of change are the same in all cases, resulting in the same slopes. For the electrical change of $K_u$ in the training process, one can turn to the voltage-controlled magnetic anisotropy (VCMA) [44-58] or the ferroelectric strain effect. In the proposed device shown in Fig. 1(a), applying a voltage across the FE layer enables the control of perpendicular anisotropy in the CoFeB/MgO stack [22]. Alternatively, one can also use a three terminal structure combining spin-orbit torque (SOT) and VCMA as shown in Fig. 8. The magnetization of the free layer can be switched by the SOT assisted by the external field $\mathbf{H}_x$. Meanwhile, the $K_u$ can be modified by applying a voltage across the MTJ stack via the VCMA effect [58]. Now that we can control the pulse width or amplitude to tune both $k$ and $c$ and we can control $K_u$ to tune $c$, the individual change of $k$ can be achieved by combining the pulse width and $K_u$, which is shown in Fig. 3(b). In addition, during the change of $K_u$, we have ensured that the free layer always maintain the easy axis along the perpendicular direction. For example, the energy barrier in Fig. 3(a) is obtained as $E_b=(K_u-0.5\mu_0 M_s^2) \times V_{FL}=15 k_B T$. The value of $K_u$ is larger than the shape anisotropy, which leads to the perpendicular magnetization. This analytical calculation is also supported by simulation, where the magnetization is tilted 45 degree away from the **z** axis and relax under $J_c=0$ to see if it can return to the **z** axis.

Fig. 4(a) shows the range of $k$ by changing both pulse width and $K_u$. When only $k$ is added into the neural network as another freedom, the training process is similar to the case with both $k$ and $c$ and a high recognition accuracy of 91.3% is achieved [see Fig. 4(b)]. In addition, it takes only 500 iterations to reach 85% accuracy, whereas it requires 800 iterations in the system without the trainable $k$. It is noted in Fig. 1(d) and Fig. 4(b) that at low iterations, the accuracy shows large difference in different systems. We attribute this to the use of different parameter initialization rules. A proper parameter initialization could lead to much improved performance at low iterations, which is discussed in detail in Appendix E. We have also repeated the simulation for 10 times with the randomly initialized weights, and the averaged accuracy and loss show negligible difference compared to that in the single run. Furthermore, we have monitored the change of $k$ during the training process and ensured that it falls in the allowed range. During this study, we find that many parameters affects the learning process, such as the learning rate. Different learning rates have been tested (0.1 and 0.01). Although there is difference in the training speed and recognition accuracy, the key result in this work is not affected, i.e., the spintronic neuron with trainable $k$ improves the performance of neural network. These parametric effects can be studied to optimize the neural network, which are out of the scope of this work. Finally, we discuss the energy consumption of the neural network. In our work, the MTJ device is used as the neuron. Its operation requires the application of a current pulse. The energy consumption is calculated as $I^2Rt$. Using an averaged $I = 15$ μA, a typical MTJ resistance $R_{MTJ} = 5$ KΩ, and an average pulse width of 15 ns, the approximated energy consumption in MTJ is $E_{\text{MTJ}}$ = 16.9 fJ per neuron. In comparison, the energy consumption of single neuron implemented by the CMOS exceeds 700 fJ [59, 60]. The energy consumption of spintronic neurons have been reported to be 0.32 fJ [11], 1 fJ [59], 18 to 36 fJ [18], and 60 fJ [60]. Therefore, the energy consumption of our proposed trainable neuron is comparable to other spintronic neurons and much smaller than the CMOS neurons. It is important to note that the updating of $k$ and $c$ in our neuron are realized by changing the pulse width, the energy of which has been included in the calculation of $E_{\text{MTJ}}$. Therefore, the proposed trainable neuron in this work can improve the performance of the neural network without introducing additional energy consumption.

**III. Conclusion**

In summary, we have proposed a spintronic neuron with trainable activation function. The sigmoid activation function with electrically controlled slope and shift is obtained through the current-induced switching of MTJ under the thermal effect. Different from previous studies, where the slope and shift are fixed during the network training, we have developed an algorithm that enable them to change dynamically during the training, resulting in faster training and a higher recognition accuracy. This trainable activation is similar to the batch normalization which is an indispensable algorithm in the deep neural network. We have also shown that the corresponding hardware implementation can be realized using a single spintronic device supplemented by the electrical control of pulse width and magnetic anisotropy. Thus, our investigation of MTJ physics to realize the trainable spintronic neuron provides insights to improve the performance of spintronic neurons, and the proposal of hardware level realization bridges the spintronic device and machine learning algorithm. These can pave the way for the physical realization of spintronic-based neural networks.

Corresponding Author: †zhuzhf@shanghaitech.edu.cn

**Acknowledgements**: This work was supported by National Key R&D Program of China (Grant No. 2022YFB4401700), Shanghai Sailing Program (Grant No. 20YF1430400) and National Science Foundation of China (Grants No. 12104301 and No. 62074099).

**APPENDIX A: Detailed explanation of the switching probability**

To illustrate the effect of pulse width on the switching probability, the switching trajectories of 100 independent runs with the same input are shown in Fig. 5(a). It can be seen from the figure that although the pulse amplitude and pulse width are exactly the same, some devices switch successfully, whereas the others return to the initial state. This is caused by the thermal noise, which is a random field and thus induces random switching. To correctly interpret this physics, the switching probability presented in this work is obtained by averaging 100 independent runs.

It is also beneficial to compare the switching current obtained in the simulation with that using the analytical solution [61]. It can be seen from Fig. 5(b) that the switching current is around 10μA. Based on the parameters used in the simulation, i.e., $\alpha = 0.0122$, $V_{FL} = 15 \times 15 \times 1.3 \times 10^{-27} m^3$, $M_s = 1.58/(4\pi) \times 10^7 A/m$, $K_u = 1.1 \times 10^6 J/m^3$, $P = 0.4$, we get $H_k = 2K_u/M_s - 4\pi M_s = 0.169T$. Therefore, the critical current $I_{crit}$ in the analytical solution is $2\alpha e M_s$

$V_{FL}H_k]/(\hbar P)$=5.8μA, which is quite close to the value obtained from the simulation. Note that this is a rough estimation. The thermal fluctuation would lead to a reduction in $I_{crit}$, whereas the use of finite pulse width in the simulation calls for a higher overdrive current. In addition, the spin torque efficiency, the shape anisotropy and the crystalline anisotropy are also changed dynamically during the magnetization switching [26].

**APPENDIX B: Definition of *k* and *c***

To illustrate the change of *k* and *c* when the input current is changed, a more detailed figure is given in Fig. 5(b). As shown in Fig. 5(b), *k* is defined as the slope at $P_{sw} = 0.5$. It can be seen that a larger pulse width of 200 ns gives a larger $k = 3.4$, compared to $k = 2.2$ at 30 ns pulse. Similarly, *c* is defined as the current shift at $P_{sw} = 0.5$ with respect to the origin, and larger pulse width results in a smaller current shift.

**APPENDIX C: Implementation of the resistive crossbar array**

An exemplary crossbar resistive array which connects the input layer and the hidden layer has been plotted in Fig. 6(b). The neurons in the input layer and hidden layer are represented as the red circles in the left-most column and bottom row, respectively. As shown in Fig. 6(a), every neuron in the input layer is connected to all the neurons in the hidden layer, which is represented by the weight $\mathbf{w}_1$ that is a 784-by-25 matrix. One can see in Fig. 6(b) that the array size of blue squares is also 784-by-25. Therefore, each blue square can be used to implement the connection between the input and hidden layer. For this purpose, one can use the resistive device such as the MRAM or RRAM as the blue square. The input to the input layer is the current, which goes through the resistive crossbar and then generates a voltage acting on the neurons in the hidden layer. In this way, the physical implementation of the neurons and synapses in the neural network can be realized.

**APPENDIX D: Derivation of the updating equations for *k* and *c***

A typical neural network consists of neural layers and their connections, corresponds to neurons and synapses in the cerebral cortex. The strength of synapses is denoted by *w*. In contrast, *k* and *c* are local parameters related to the neurons. During the training of neural network, *w* and *k(c)* are changed after each iteration so that the total loss is reduced. The equation guides the change of *w* is derived based on the principle of gradient descent and backpropagation. Since *k* and *c* proposed in this work are physically different from *w*, their updating equations are also different. Therefore, we have to derive the backpropagation equation for the training of *k* and *c*.

The neural network is shown in Fig. 6(a). The derivation flowchart is shown in Fig. 7. Some notations are defined below:

$n^{[l]}$ – The number of neurons in the layer *l*.

$a^{[l]}$ – The vector of activations of the neurons in the layer *l*, of the size ($n^{[l]} \times 1$).

$z^{[l]}$ – The vector of the weighted output of the neurons in the layer *l*, of the size ($n^{[l]} \times 1$).

$w^{[l]}$ – The weight matrix associated with the layers *l* and *l*-1, of the size ($n^{[l]} \times n^{[l-1]}$).

$g^{[l]}$ – The activation function applied to the output of the neurons in the layer *l*, $a^{[l+1]} = g^{[l]}(z^{[l+1]})$.

In the input layer, $a^{[1]} = X$.

In the hidden layer, $z^{[2]} = w^{[1]} \cdot a^{[1]}$ and $a^{[2]} = g^{[2]}(k^{[2]} \cdot (z^{[2]} + c^{[2]}))$ where $z^{[2]}$ represents the summation of weighted input with weights $w^{[1]}$, $g^{[2]}$ represents the activation function of neuron with parameters $k^{[2]}$ and $c^{[2]}$. $a^{[2]}$ represents the output of neuron.

In the output layer, $z^{[3]} = w^{[2]} \cdot a^{[2]}$ and $a^{[3]} = g^{[3]}(k^{[3]} \cdot (z^{[3]} + c^{[3]}))$ where $z^{[3]}$ represents the summation of weighted input with weights $w^{[2]}$, $g^{[3]}$ represents the activation function of neuron with parameters $k^{[3]}$ and $c^{[3]}$. $a^{[3]}$ represents the output of neuron.

The neural network then gives the predicted value $h_\theta = a^{[3]}$, which is used as input to calculate the cost function $J = -[y \log h_\theta + (1 - y) \log(1 - h_\theta)]$ that represents the difference between $h_\theta$ and the true value $y$.

The target of neural network training is to minimize the cost function by finding the optimum value of $w^{[1]}$, $w^{[2]}$, $k^{[2]}$, $k^{[3]}$, $c^{[2]}$ and $c^{[3]}$. The updating equation of these parameters are based on the principle of gradient descent shown as Eq. 1:

$$\begin{cases} w^{[2]} = w^{[2]} - \alpha \frac{\partial J}{\partial w^{[2]}} \\ k^{[3]} = k^{[3]} - \alpha \frac{\partial J}{\partial k^{[3]}} \\ c^{[3]} = c^{[3]} - \alpha \frac{\partial J}{\partial c^{[3]}} \\ w^{[1]} = w^{[1]} - \alpha \frac{\partial J}{\partial w^{[1]}} \\ k^{[2]} = k^{[2]} - \alpha \frac{\partial J}{\partial k^{[2]}} \\ c^{[2]} = c^{[2]} - \alpha \frac{\partial J}{\partial c^{[2]}} \end{cases} \quad \text{(Eq. 1)}$$

The derivative of $J$ can be obtained based on the backpropagation. Firstly, we can calculate

$$\frac{\partial J}{\partial k^{[3]}} = \frac{\partial J}{\partial a^{[3]}} \cdot \frac{\partial a^{[3]}}{\partial k^{[3]}} = -(\frac{y}{a^{[3]}} - \frac{1-y}{1-a^{[3]}}) \cdot \frac{\partial a^{[3]}}{\partial k^{[3]}} = ((-(\frac{y}{a^{[3]}} - \frac{1-y}{1-a^{[3]}})) \cdot *\sigma'(k^{[3]} \cdot (z^{[3]} + c^{[3]}))) \cdot *z^{[3]}$$

$$\frac{\partial J}{\partial c^{[3]}} = \frac{\partial J}{\partial a^{[3]}} \cdot \frac{\partial a^{[3]}}{\partial c^{[3]}} = -(\frac{y}{a^{[3]}} - \frac{1-y}{1-a^{[3]}}) \cdot \frac{\partial a^{[3]}}{\partial c^{[3]}} = (-(\frac{y}{a^{[3]}} - \frac{1-y}{1-a^{[3]}})) \cdot *\sigma'(k^{[3]} \cdot (z^{[3]} + c^{[3]}))$$

$$\frac{\partial J}{\partial w^{[2]}} = \frac{\partial J}{\partial z^{[3]}} \cdot a^{[2]T} = \frac{\partial J}{\partial a^{[3]}} \cdot \frac{\partial a^{[3]}}{\partial z^{[3]}} \cdot a^{[2]T} = -(\frac{y}{a^{[3]}} - \frac{1-y}{1-a^{[3]}}) \cdot \frac{\partial a^{[3]}}{\partial z^{[3]}} \cdot a^{[2]T}$$
$$= (((-(\frac{y}{a^{[3]}} - \frac{1-y}{1-a^{[3]}})) \cdot *\sigma'(k^{[3]} \cdot (z^{[3]} + c^{[3]}))) \cdot *k^{[3]}) \cdot a^{[2]T}$$

Based on $\frac{\partial J}{\partial w^{[2]}}$, we can get

$$\frac{\partial J}{\partial w^{[1]}} = \frac{\partial J}{\partial z^{[2]}} \cdot a^{[1]T} = \frac{\partial J}{\partial z^{[2]}} \cdot X^T = \frac{\partial J}{\partial a^{[2]}} \cdot \frac{\partial a^{[2]}}{\partial z^{[2]}} \cdot X^T$$
$$= w^{[2]T} \cdot \frac{\partial J}{\partial z^{[3]}} \cdot \frac{\partial a^{[2]}}{\partial z^{[2]}} \cdot X^T = ((w^{[2]T} \cdot \frac{\partial J}{\partial z^{[3]}} \cdot *\sigma'(k^{[2]} \cdot (z^{[2]} + c^{[2]}))) \cdot *k^{[1]}) \cdot X^T$$
$$= (((w^{[2]T} \cdot (((-(\frac{y}{a^{[3]}} - \frac{1-y}{1-a^{[3]}})) \cdot *\sigma'(k^{[3]} \cdot (z^{[3]} + c^{[3]}))) \cdot *k^{[3]})) \cdot *\sigma'(k^{[2]} \cdot (z^{[2]} + c^{[2]}))) \cdot *k^{[2]}) \cdot X^T$$

$$\frac{\partial J}{\partial k^{[2]}} = \frac{\partial J}{\partial a^{[2]}} \cdot \frac{\partial a^{[2]}}{\partial k^{[2]}} = w^{[2]T} \cdot \frac{\partial J}{\partial z^{[3]}} \cdot \frac{\partial a^{[2]}}{\partial k^{[2]}} = (w^{[2]T} \cdot \frac{\partial J}{\partial z^{[3]}} \cdot *\sigma'(k^{[2]} \cdot (z^{[2]} + c^{[2]}))) \cdot *z^{[2]}$$
$$= ((w^{[2]T} \cdot (((-(\frac{y}{a^{[3]}} - \frac{1-y}{1-a^{[3]}})) \cdot *\sigma'(k^{[3]} \cdot (z^{[3]} + c^{[3]}))) \cdot *k^{[3]})) \cdot *\sigma'(k^{[2]} \cdot (z^{[2]} + c^{[2]}))) \cdot *z^{[2]}$$

$$\frac{\partial J}{\partial c^{[2]}} = \frac{\partial J}{\partial a^{[2]}} \cdot \frac{\partial a^{[2]}}{\partial c^{[2]}} = w^{[2]T} \cdot \frac{\partial J}{\partial z^{[3]}} \cdot \frac{\partial a^{[2]}}{\partial c^{[2]}} = w^{[2]T} \cdot \frac{\partial J}{\partial z^{[3]}} \cdot *\sigma'(k^{[2]} \cdot (z^{[2]} + c^{[2]}))$$
$$= (w^{[2]T} \cdot (((-(\frac{y}{a^{[3]}} - \frac{1-y}{1-a^{[3]}})) \cdot *\sigma'(k^{[3]} \cdot (z^{[3]} + c^{[3]}))) \cdot *k^{[3]})) \cdot *\sigma'(k^{[2]} \cdot (z^{[2]} + c^{[2]}))$$

These above equations are then used in Eq. 1 to guide the change of $w^{[1]}$, $w^{[2]}$, $k^{[2]}$, $k^{[3]}$, $c^{[2]}$ and $c^{[3]}$. After several iterations, the minimum $J$ can be obtained.

**APPENDIX E: Initialization rule for *k*, *c*, and *w***

The results shown in Fig. 1(d) is obtained by averaging 10 independent runs. The parameters in each run are randomly initialized. In our code, we initialize the parameter follows Ref. [62], i.e., the initial **w** is randomly distributed between $[-\varepsilon_{init}, \varepsilon_{init}]$ with

$\varepsilon_{init} = \sqrt{6} / \sqrt{n_{in} + n_{out}}$ where $n_{in}$ and $n_{out}$ denote the number of neurons at the input and output of the synapse, respectively. This has also been implemented to $\mathbf{k}_1$, $\mathbf{k}_2$, $\mathbf{c}_1$, $\mathbf{c}_2$ in our code. This leads to the results shown in Fig. 1(d) that the two curves show large difference at low iterations. By examining the raw data, we find that, at low iterations, the accuracy in the system with trainable *k&c* is always lower than the system without *k&c*.

However, as we have noted in the paper as well as the learned value of **k** and **c** shown in Fig. 2(a), **k** and **c** are fundamentally different from **w**. Therefore, it is reasonable to use a different initialization rule for **k** and **c**. As we have shown in Fig. 2(a) that the learned **k** ranges from −4 to 4 and **c** ranges from −1 to 1, we propose to set $\varepsilon_{init,k}$ and $\varepsilon_{init,c}$ to [−4, 4] and [−1, 1], respectively. The corresponding training shows significant improvements in the accuracy at low iterations, which is shown in Fig. 9. Similarly, we have checked that the accuracy in every run is better than that in the system without *k&c*. The difference in these results call for more theoretical studies on the initialization rule of nonlocal parameters such as *k* and *c* presented in this work.

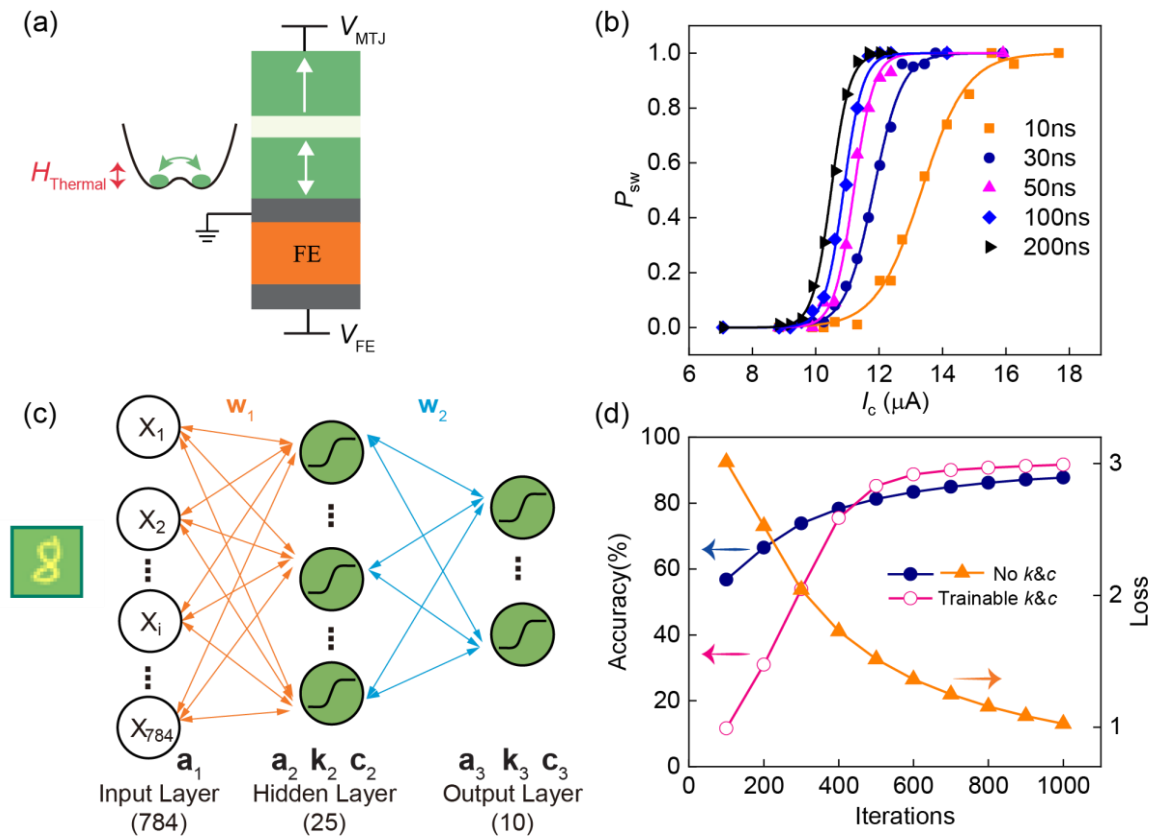

**Figure 1.** (a) Illustration of the device structure. (b) $P_{sw}$ as a function of $I_c$ under different pulse

widths in the system with $E_B=7.5k_BT$. The simulation results and fitting are denoted by the symbols and lines, respectively. (c) A three-layer neural network. $\mathbf{w}_1$ and $\mathbf{w}_2$ denote the resistive connections of the neurons between different layers. The subscript 2 and 3 of **a**, **k** and **c** refer to the input layer, the hidden layer and the output layer, respectively. (d) The evolution of loss and accuracy in the system without trainable $k$ and $c$ are denoted by the filled symbols. The evolution of accuracy in the system with trainable $k$ and $c$ is denoted by the filled circles. The source code corresponding to the trainable $k$&$c$ can be accessed at https://github.com/zhuzibn/trainable_activation_function/tree/main/2022TrainableNeuon.

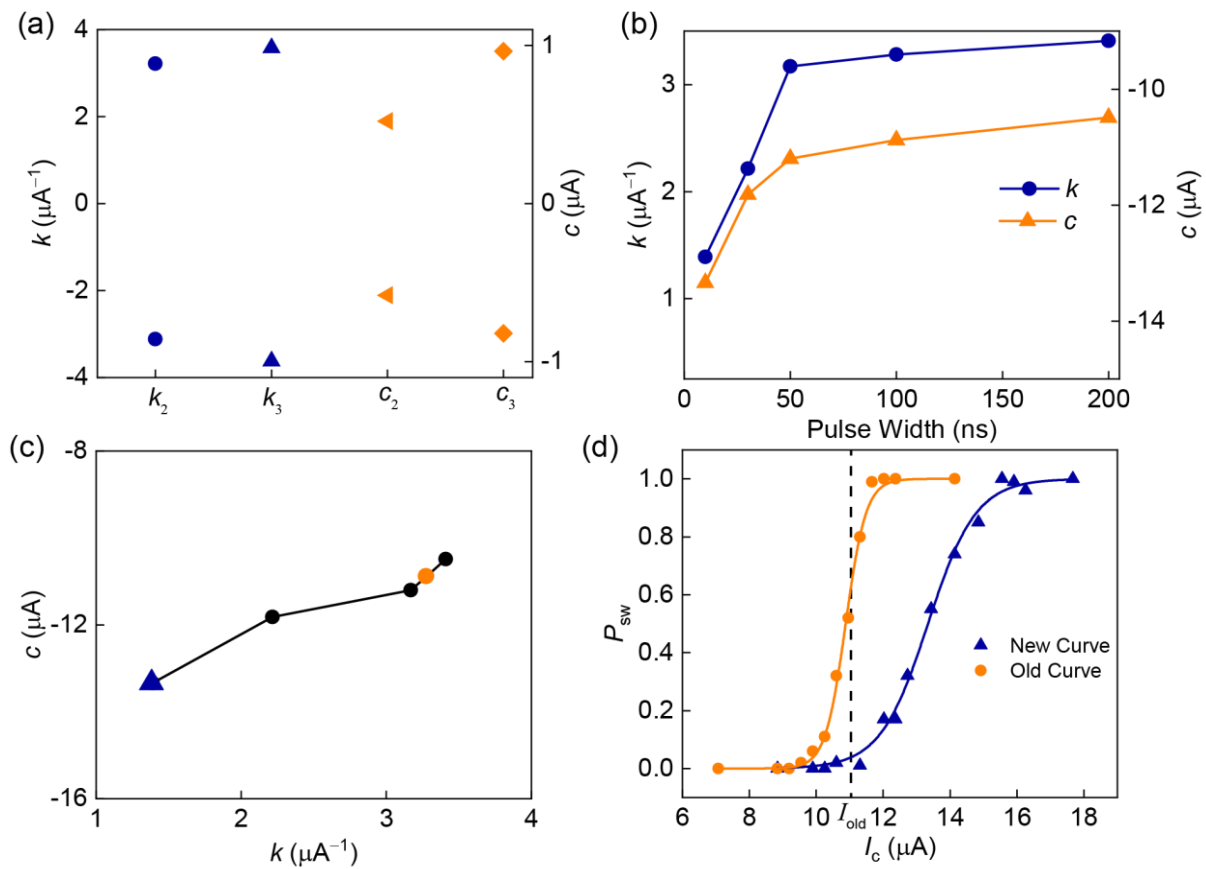

**Figure 2.** (a) Maximum ranges of $k_2$, $k_3$, $c_2$ and $c_3$ extracted from the software training. (b) Physically allowed $k$ and $c$ under different pulse widths. The values are extracted from Fig.

1(b). (c) Relation between $k$ and $c$ obtained from Fig. 2(b). (d) Activation functions corresponding the two points marked in Fig. 2(c).

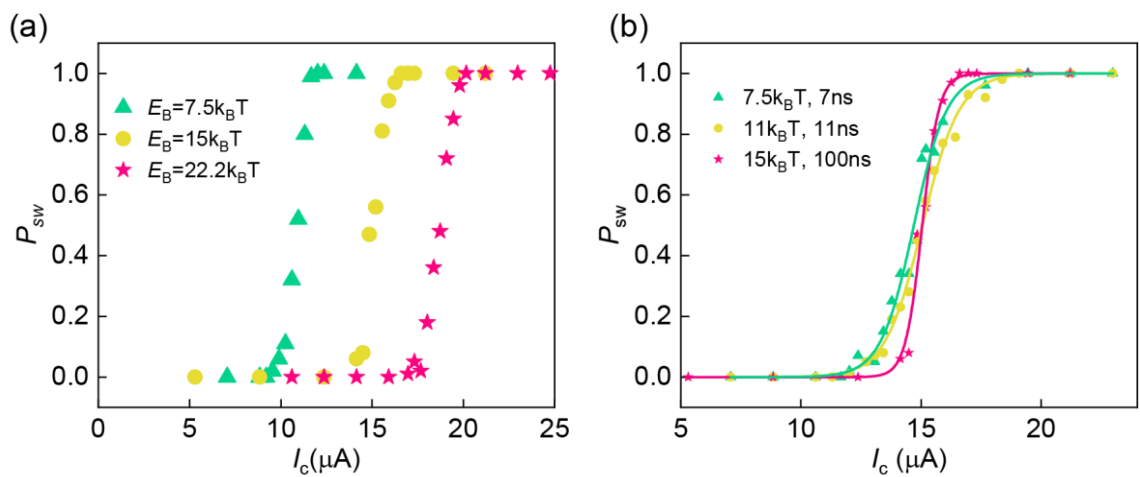

**Figure 3.** (a) $P_{sw}$ as a function of $I_c$ under different $E_B$ which is obtained by changing $K_u$ that $E_B=(K_u-0.5\mu_0 M_s^2)\times V_{FL}$. The source code corresponding to the system with $15k_BT$ under 30ns pulse and $I_c$=15.6 µA can be accessed at https://github.com/zhuzibn/ODE_LLG_integral/tree/master/2022TrainableNeuron. (b) $P_{sw}$ as a function of $I_c$ under different combination of $E_B$ and pulse widths.

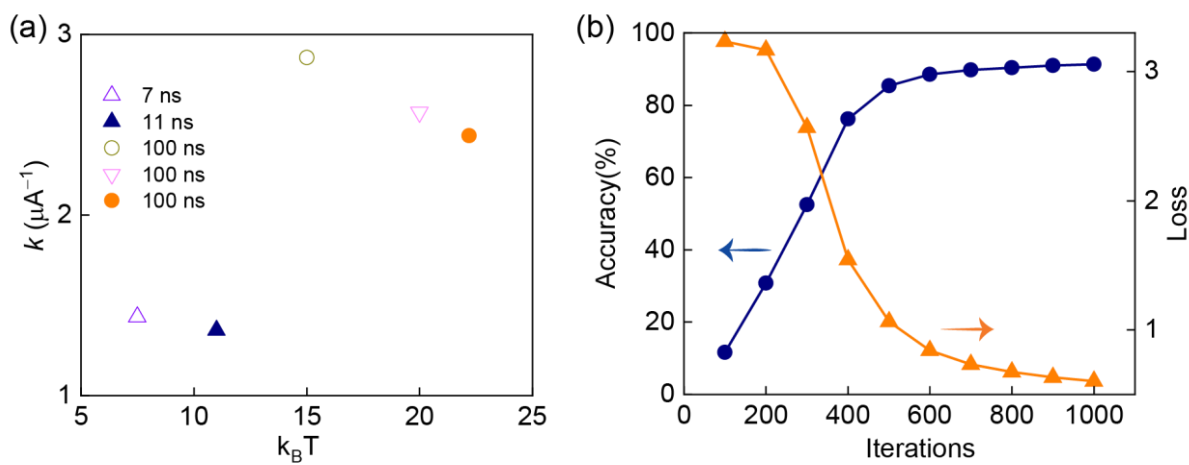

**Figure 4.** (a) $k$ under different combination of $E_B$ and pulse widths. (b) The evolution of loss

and accuracy in the system with trainable $k$.

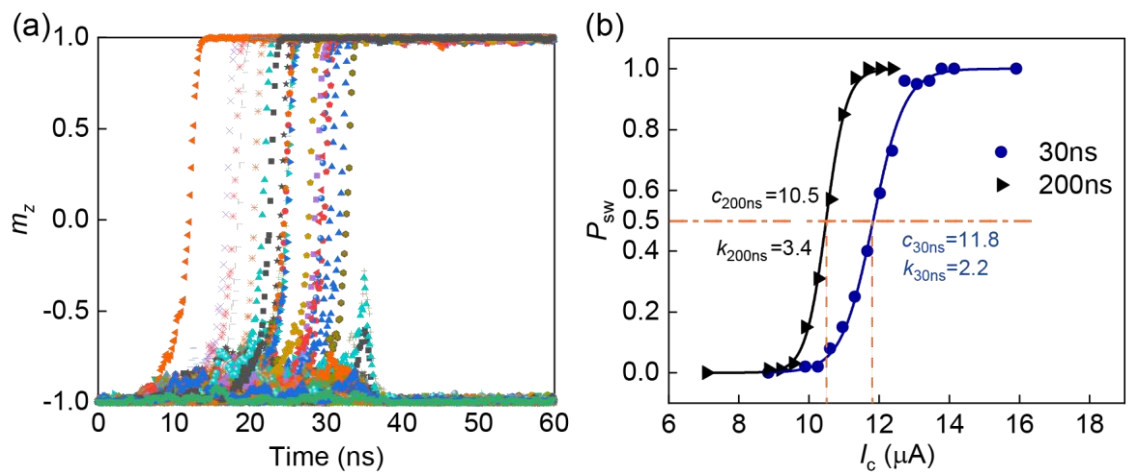

**Figure 5.** (a) The switching trajectories of 100 independent runs at $I_c = 11.3$ μA and 30 ns pulse width. The initial state is $m_z = -1$. The final state with $m_z = 1$ indicates successful magnetization switching. (b) Illustration of $k$ and $c$ at different pulse widths.

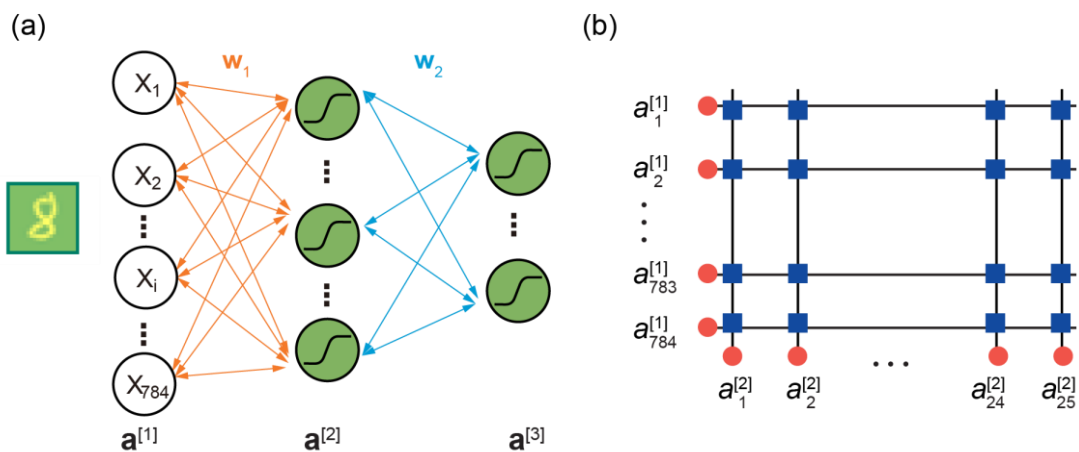

**Figure 6.** (a) Structure of a three-layer neural network. (b) Illustration of the crossbar resistive array between the input layer and the hidden layer, where the resistance of blue squares represents the weights and the red circles denote the neurons.

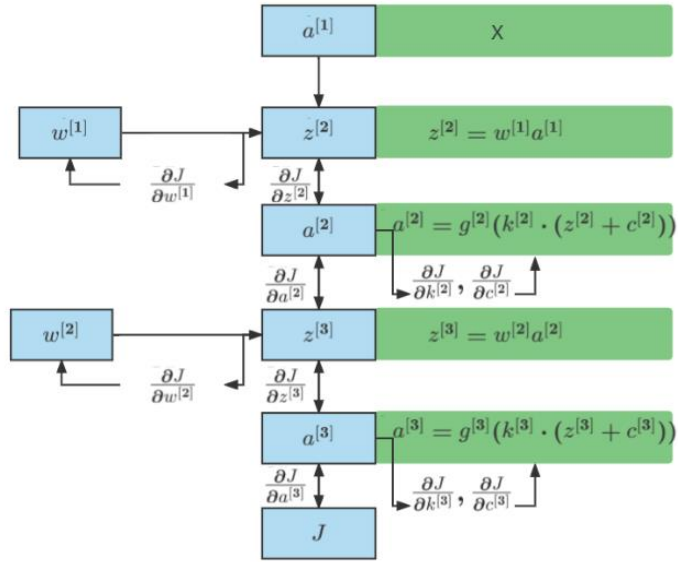

**Figure 7.** A flow chart describing the calculation performed in the neural network with trainable $k$ and $c$.

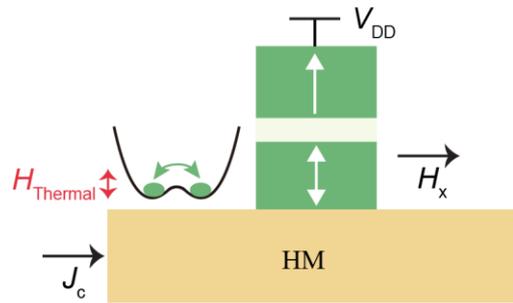

**Figure 8.** Illustration of the three terminal device combining SOT and VCMA.

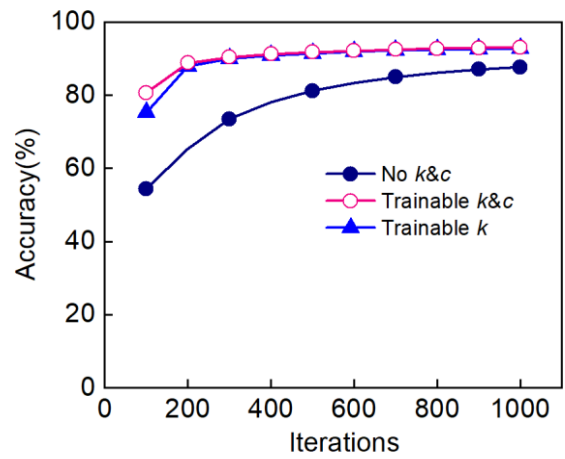

**Figure 9.** The evolution of accuracy after changing $\varepsilon_{init,k}$ and $\varepsilon_{init,c}$ to $[-4, 4]$ and $[-1, 1]$, respectively. Each point is obtained by averaging the results in 10 independent runs.